\documentclass[iop]{emulateapj}

\slugcomment{version Aug 29, 2012}

\shorttitle{Power spectrum analysis of the solar surface}
\shortauthors{Katsukawa \& Orozco Su\'arez}

\begin{document}

\title{Power spectra of velocities and magnetic fields on the solar surface
and their dependence on the unsigned magnetic flux density}

\author{Y. Katsukawa\altaffilmark{1} and D. Orozco Su\'arez\altaffilmark{2,3}}
\affil{
$^{1}$National Astronomical Observatory of Japan, 2-21-1 Osawa, Mitaka, Tokyo
181-8588, Japan; yukio.katsukawa@nao.ac.jp\\
$^{2}$Instituto de Astrof\'{i}sica de Canarias, E-38205 La Laguna, Tenerife, Spain}
\altaffiltext{3}{Dept. Astrof\'isica, Universidad de La Laguna, E-38206 La Laguna,
Tenerife, Spain}
%%\email{yukio.katsukawa@nao.ac.jp}

\begin{abstract}
We have performed power spectral analysis of surface temperatures, velocities, and 
magnetic fields, using spectro-polarimetric data taken with the Hinode Solar Optical 
Telescope. When we make power spectra in a field-of-view covering the super-granular 
scale, kinetic and thermal power spectra have a prominent peak at the granular scale 
while the magnetic power spectra have a broadly distributed power over various spatial 
scales with weak peaks at both the granular and supergranular scales. To study the power 
spectra separately in internetwork and network regions, power spectra are derived 
in small sub-regions extracted from the field-of-view. We examine slopes of the power 
spectra using power-law indices, and compare them with the unsigned magnetic flux 
density averaged in the sub-regions. The thermal and kinetic spectra are 
steeper than the magnetic ones at the sub-granular scale in the internetwork
regions, and the power-law indices differ by about 2. The power-law indices of the 
magnetic power spectra are close to or smaller than -1 at that scale, which suggests 
the total magnetic energy mainly comes from either the granular scale magnetic 
structures or both the granular scale and smaller ones contributing evenly. 
The slopes of the thermal and kinetic power spectra become less steep with increasing 
unsigned flux density in the network regions. The power-law indices of all the
thermal, kinetic, and magnetic power spectra become similar when the unsigned 
flux density is larger than 200~Mx~cm$^{-2}$. 
\end{abstract}


\section{Introduction}

On the solar surface, interaction between magnetic fields and surface convection 
produces varieties of structures over the broad spatial scale from $10^5$~km, 
such as sunspots and active regions, down to $\sim 100$~km or even below, such 
as small flux concentrations. Spatial power spectra of velocity and magnetic fields on 
the solar surface provide a clue for understanding at which scale kinetic 
and magnetic energies are injected, transferred, and dissipated on the solar surface
\cite[see recent reviews by][]{2010LRSP....7....2R, 2012ASPC..455...17A}. 
Basic information of the granular convection was retrieved from intensity and 
Doppler-shift images showing temperature and velocity fluctuations on the 
solar surface. Their power spectra were analyzed by various authors using high
resolution observations taken with ground-based telescopes 
\citep{1955ApJ...121..216F,1969SoPh....9...39B,1971SoPh...19..297R,
1982A&A...111..272D, 1989ssg..conf..101M,1991A&A...248..245R,1993A&A...271..589E}. 
They discussed how the turbulent convection on the solar surface makes slopes 
of the power spectra toward a high wavenumber domain by comparing them with the 
Kolmogorov's $k^{-5/3}$ power-law, where $k$ is the wavenumber. Power spectra 
of surface magnetic fields were also analyzed using longitudinal magnetograms 
taken with ground-based telescopes and SOHO/MDI \citep{1973ApJ...179..949N,
1974ApJ...190..441N,1981ApJ...247..300K, 1997SoPh..171..269L,
2001SSRv...95....9P,2001SoPh..201..225A}. But it has been hard to study the 
power spectra with enough confidence at the granular and sub-granular scales 
because of the lack of spatial resolution in polarimetric observations.

In most studies using ground-based telescopes, power spectra were
significantly affected not only by the instrumental resolving performance 
but also degradation of the image quality due to atmospheric seeing, especially 
when we study the power spectra at the scale smaller than granules though 
calibration of such degradation was attempted \citep[e.g.][]{1997SoPh..171..269L}. 
It is also important to distinguish between different observed regions 
when we make power spectra, because unsigned magnetic flux densities have 
significant variation depending on internetwork, network, and active regions. 
Influence of magnetic fields to the surface convection cannot be negligible 
in strong field regions while it may be small in weak field regions. 
\cite{2005ApJ...629.1141A} and \cite{2010ApJ...720..717A} studied magnetic
power spectra in various active regions, and suggested correlation between
the power-law index of the magnetic spectrum and flare productivity of an 
active region. But it has not been understood yet how the power spectra 
changes from internetwork regions to network ones.

The Solar Optical Telescope \cite[SOT][]{2008SoPh..249..167T,2008SoPh..249..197S, 
2008SoPh..249..233I,2008SoPh..249..221S} aboard Hinode \citep{2007SoPh..243....3K}
is a suitable instrument for studying the behavior of the power spectra at the 
granular scale and even below owing to the diffraction-limited performance 
of the 50~cm aperture and stable image quality. \cite{2010A&A...512A...4R}
examined power spectra of intensities, transverse and longitudinal velocities
using filtergram data taken with the Hinode SOT. We use two-dimensional maps taken 
with the spectro-polarimeter (SP) of SOT, which not only provides intensity and 
velocity maps, but also magnetic field maps with the same instrument. 
It is becoming important to understand properties of small-scale magnetic 
structures in internetwork regions because the Hinode SP observations 
suggested that small-scale fields are possibly generated and sustained by 
turbulent small-scale dynamo \citep[e.g.][]{2009A&A...495..607I, 
2009ApJ...693.1728P,2010A&A...513A...1D, 2011ApJ...737...52L}. The most common
indicator is the probability distribution functions (PDFs) of magnetic field 
strength or magnetic flux. \cite{2012A&A...541A..17S} and 
\cite{2012ASPC..455...17A} used the magnetic power spectra for investigating 
the small-scale magnetic structures using the Hinode SP data.

In this article, we make power spectrum analyses of the surface temperatures,
velocities, and magnetic fields using the Hinode SP data, including calibration 
of the instrumental resolution performance. Section 2 describes how the SP 
data are processed to make the power spectra. Section 3 presents the
power spectra covering both the supergranular and granular scales. In Section 4, 
the power spectra are created in small regions to distinguish internetwork and 
network regions. We study how the power spectra change with the unsigned 
magnetic flux density.

\section{Data analysis}
\subsection{Data sets and reduction}

\begin{figure}[t]
\plotone{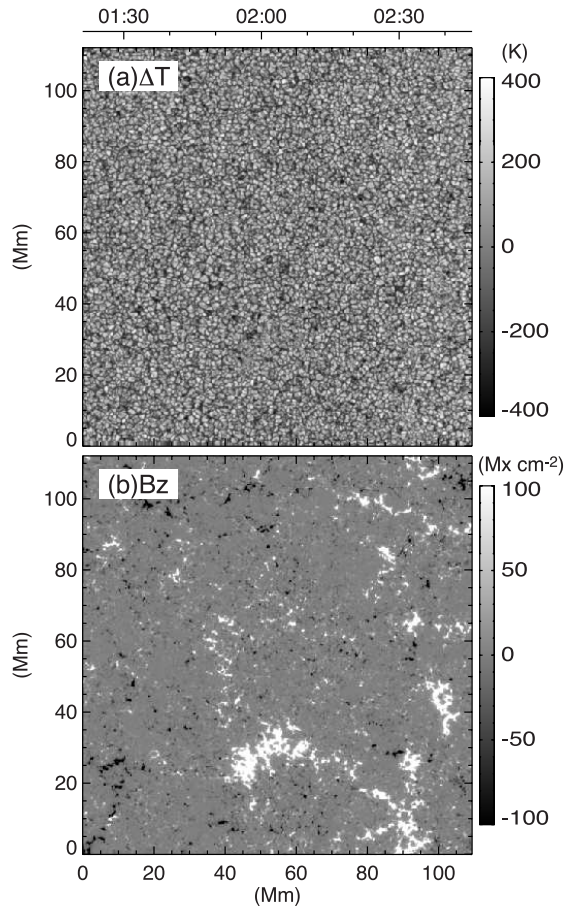}
\caption{Two dimensional maps of (a) the temperature fluctuation $\Delta T$
and (b) the vertical component of magnetic flux densities $B_z$ obtained 
with the Hinode SP on 24 April 2007, as an example of the data set used in 
the power spectrum analysis. The FOV is 1000 steps in the scanning direction
and 1024 pixels along the slit, which covers a square of about 110~Mm. The 
hours of the SP scanning observation are shown at the top of the figure.
\label{f0}} 
\end{figure}

We use a data set taken with the SP of the Hinode SOT.  The data set consists 
of 30 SP scans taken between November 2006 and December 2007. The SP recorded 
full Stokes spectra of the two Fe I lines at 630.15~nm\ and 630.25~nm\ with 
the normal map mode whose spatial sampling was about 0.15'' per pixel and 
integration time is 4.8 sec per slit position. The field-of-views (FOVs) of 
the scans were equal to or more than 1000 steps in the scanning direction and 
1024 pixels along the slit, and containing the solar disk center. It 
took about 85 minutes to complete the scan of each FOV. Their observing 
targets were mostly in quiet regions, but some of the scans included active 
regions also. We identify 23 scans out of 30 containing only the quiet regions 
using the condition in which an averaged density of unsigned flux 
$\left<|Bz|\right>$ in the FOV is smaller than 20~Mx~cm$^{-2}$. The 
SP data are calibrated with the standard routine {\it sp\_prep} available 
under the Solar SoftWare (SSW). The wavelength positions of the two Fe I 
lines are calibrated with an average line-center position outside of a sunspot.

The SP data provide two-dimensional maps of surface temperatures, line-of-
sight (LOS) velocities, and magnetic field vectors. The temperatures $T$ 
are obtained using continuum intensities averaged in the spectral 
window outside of the Fe I lines and under assumption of black body 
radiations. We assume that average temperature $T_0$ is 6520~K\ outside 
of sunspots to convert the observed continuum intensities to the 
temperature. The LOS velocities $v_z$ are obtained using Doppler shifts 
of the Fe I line at 630.15~nm. The magnetic field vectors 
$(B_x, B_y, B_z)$ are determined from wavelength-integrated Zeeman-induced
polarization signals with the method described in \cite{2008ApJ...672.1237L}.
The x, y, and z components correspond to ones in the scanning, slit, and 
line-of-sight directions, respectively. It is not necessary to convert
the observed vectors into the ones in the local reference frame because
the maps were taken near the disk center. To derive the transverse component 
$(B_x, B_y)$, the 180 degrees ambiguity in the azimuth angle has to be resolved. 
This is simply done by choosing the azimuth to be closest to the computed 
potential field using $B_z$ as an input. Fig. \ref{f0} shows two dimensional 
maps of the temperature fluctuation $\Delta T=T-T_0$ and the vertical component 
of magnetic flux densities $B_z$ as an example of the data set.

\subsection{Derivation of power spectra}
For deriving spatial power spectra, we apply the two-dimensional Fourier-transform
to each map of the above parameters, 
\begin{equation}
P(f;k_x,k_y) = \left|\frac{1}{N^2}\sum_{x,y} f(x, y)\exp\left(-2\pi i(k_xx+k_yy)\right)\right|^2 ,
\end{equation}
where $f(x,y)$ is either $\Delta T=T-T_0$, $v_z$, $B_x$, $B_y$, or 
$B_z$, and $N$ the number of pixels in one direction and 1024 in our analysis.
The wavenumbers in the x and y directions are represented as $k_x$ and 
$k_y$.\footnote{The definition of the wavenumber is the same as the one in 
\cite{2010A&A...512A...4R}, in which the spatial scale corresponding to 
the wavenumber is provided by the inverse of the wavenumber. Some other papers
\citep[e.g.][]{1997SoPh..171..269L,2001SoPh..201..225A} used 
$\exp\left(-i(k_xx+k_yy)\right)$ in the Fourier transform, in which the wavenumber is 
a factor of $2\pi$ larger than the one used in this article, and the spatial scale 
is provided by inverse of the wavenumber multiplied by $2\pi$.} To have 1024 pixels 
in both the x(scanning)- and y(slit)- directions for simplifying the Fourier-transform 
in Eq. (1), we extract 1024 steps closest to the disk center when the number 
of steps in a scan is more than 1024. Because our selection criteria of the SP 
data set is that the number of pixels in the x-direction is more than 1000, the 
number of pixels in the x-direction is fewer than 1024 pixels in some of the maps. 
For those maps, we apply padding at both the ends in the x-direction to make a 
map consisting of $1024\times 1024$ pixels. We confirmed that the influence of 
the edges and padding is negligible when we apply a Hamming window in the 
Fourier-transform.

\begin{figure}[t]
%\epsscale{0.6}
\plotone{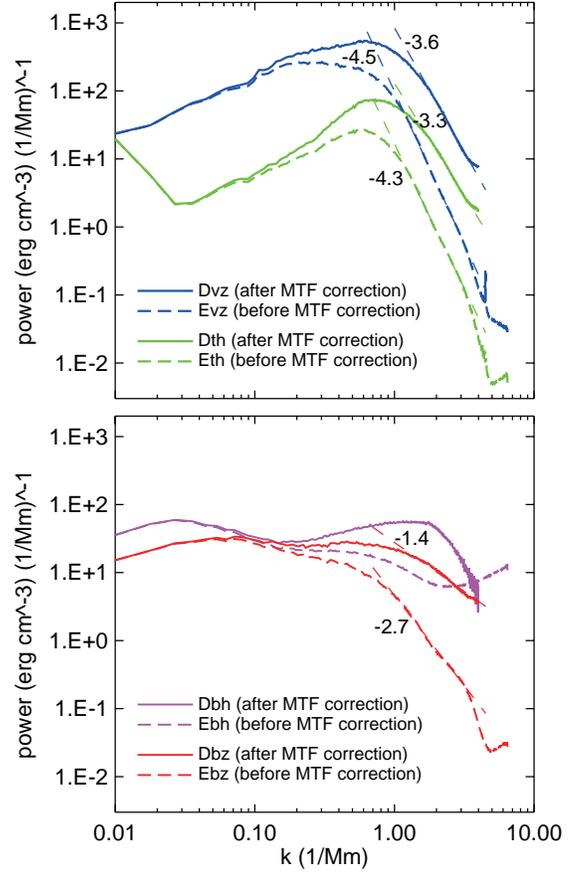}
\caption{Spatial power spectra of temperatures (green curves in the top panel), 
LOS velocities (blue curves in the top panel), and horizontal and
vertical magnetic fields (purple and red curves in the bottom panel). The 
dashed and solid curves indicate power spectra before and after the instrument
MTF and noise correction, respectively. The numbers indicate the power-law
indices $\alpha$ obtained in the wavenumber range between $k=$ 1.5 and 
3.5~Mm$^{-1}$.
\label{f1}}
\end{figure}

The two dimensional spatial power spectra $P(f;k_x,k_y)$ is then converted 
to their one dimensional form $P(f;k)$ by integrating the two-dimensional 
power spectrum in azimuthal angle of the wavenumber $(k_x, k_y)$,
\begin{equation}
P(f;k)=2\pi k \left<P(f;k_x,k_y)\right>_{k' \in [k-\Delta k/2,k+\Delta k/2)} ,
\end{equation}
where $k'=\sqrt{k_x^2+k_y^2}$, and $\Delta k$ is the sampling interval of 
the wavenumber, $\Delta k=1/(N\Delta x)$, where $\Delta x$ is the 
pixel scale and 0.15'' or 0.11~Mm on the solar surface. The power spectra thus 
derived are normalized using the following equations to represent them in a 
unit of energy density per unit wavenumber, erg cm$^{-3}$ (1/Mm)$^{-1}$, 
\begin{eqnarray}
E_{th}(k)&=&\frac{3}{2}\frac{nk_{B}}{T_{0}}\frac{P(\Delta T;k)}{\Delta k}\\
E_{vz}(k)&=&\frac{1}{2}\rho\frac{P(v_z;k)}{\Delta k}\\
E_{bh}(k)&=&\frac{1}{8\pi}\frac{\left(P(B_{x};k)+P(B_{y};k)\right)}{\Delta k}\\
E_{bz}(k)&=&\frac{1}{8\pi}\frac{P(B_{z};k)}{\Delta k},
\end{eqnarray}
respectively, where $k_{B}$ is the Boltzman constant, $n$ the number 
density on the solar surface, where we assume $n=1\times 10^{17}$~cm$^{-3}$
as a typical photospheric number density, $\rho$ the mass density on the 
solar surface, where we assume $\rho=2.7\times 10^{-7}$~g~cm$^{-3}$. 
The method used here is essentially consistent with the ones used in 
previous studies \citep[such as][]{2010A&A...512A...4R} except the 
normalization. The power spectra $E_{th}(k)$, $E_{vz}(k)$, $E_{bh}(k)$, 
and $E_{bz}(k)$ are created from each map. The power spectra averaged in 
the quiet regions (i. e. using 23 maps out of 30) are shown as dashed 
curves in Fig. \ref{f1}. 

\begin{figure}[t]
%\epsscale{0.8}
\plotone{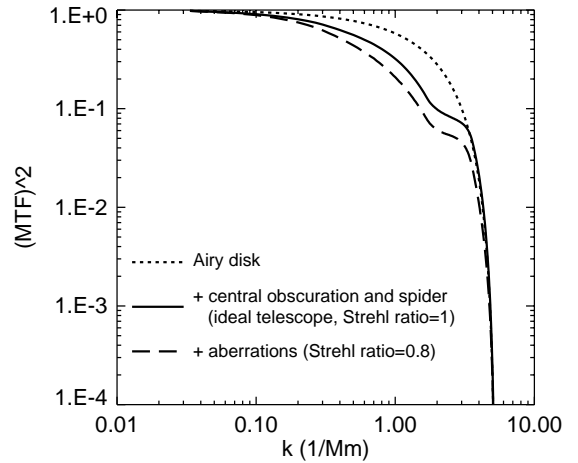}
\caption{Square of the Modulation Transfer Functions (MTFs) of the instrument
at 630~nm. The dotted curve represents a square of MTF of an Airy disk of a 
50 cm diameter aperture. The solid curve is for an ideal telescope without 
any aberrations (i.e. the Strehl ratio is unity), which includes the 
central obscuration and spiders. The dashed curve is for a telescope with 
some aberrations whose Strehl ratio is about 0.8.
\label{f2}}
\end{figure}

Because the power in the higher wavenumber domain is affected by the 
instrumental spatial resolution, it is necessary to calibrate the 
optical transfer function to obtain the shape of the power spectra on
the solar surface. The modulation transfer function (MTF) of the 
instrument is evaluated at 630~nm and is shown in Fig. \ref{f2}. 
Though the telescope achieved diffraction limited performance 
\citep{2008SoPh..249..197S,2008A&A...487..399W}, there are some 
residual aberrations in the optical performance. In this analysis, 
we use the MTF when the telescope has some aberration whose Strehl 
ratio is 0.8 (dashed curve in Fig. \ref{f2}). \footnote{ A 
wavefront error of the Hinode SOT was evaluated at G-band 430.5~nm 
with the Broadband Filter Imager (BFI) using the phase diversity 
method where on-focus and out of focus images were taken sequentially
by moving a focusing lens (Y. Suematsu 2012, in preparation). The MTF 
used in this study is derived on the assumption that the wavefront 
error derived in BFI can be applied to SP whose observing wavelength is 
630~nm. } The MTF is similar to the one used in 
\cite{2008A&A...484L..17D} who studied contrast of granules with 
Hinode SP data. Though the MTF falls down to zero at around 
$k=5$~Mm$^{-1}$ (corresponding to 200~km) as is shown in 
Fig. \ref{f2}, the observed power spectra shown as dashed curves
in Fig. \ref{f1} have significant power in the wavenumber range 
higher than 5~Mm$^{-1}$. The residual power in the high 
wavenumber range is mainly attributed to noises in the observed 
maps. The power due to the noises should be subtracted for 
calibrating the MTF, otherwise higher power appears erroneously 
in that range. When the data has a spatially uncorrelated white 
noise in a two-dimensional map, the noise spectrum $N_{f}(k)$ is 
expected to be proportional to $k$ in one-dimensional power spectra
\citep[see][]{2010A&A...512A...4R}. Thus the noise power $N_{f}(k)$ 
of each parameter is estimated using the linear regression of the 
power spectrum $E_{f}(k)$ in the wavenumber range between $k=$ 5 
and 6~Mm$^{-1}$. MTF-calibrated power spectra $D_f(k)$ are then
 obtained using the following equation \citep[see][]{1997SoPh..171..269L};
\begin{equation}
D_f(k)=(E_f(k)-N_{f}(k))/MTF(k)^2.
\end{equation}
The calibrated power spectra averaged in the quiet regions are shown 
as solid curves in Fig. \ref{f1}, which show that the power in the 
wavenumber range higher than $k=0.1$~Mm$^{-1}$ is recovered by the 
calibration of the instrument MTF. The power in the wavenumber range 
higher than 4~Mm$^{-1}$ is acknowledged to be sensitive to how 
we subtract the noise power. Thus, the calibrated power spectra 
$D_f(k)$ are shown up to 4~Mm$^{-1}$ in Fig.~\ref{f1}. 

\section{Power spectra at the granular and supergranular scales}

The SP FOVs used to make the power spectra are a square of about
110~Mm, which contain several supergranules whose typical diameters 
are 20 -- 30~Mm, and make it possible to cover both the granular
and supergranular scales. The thermal power spectrum $D_{th}(k)$ 
and the velocity power spectrum $D_{vz}(k)$ averaged in the quiet 
regions look very similar, as is shown in the top panel of 
Fig.~\ref{f1}. This can be naturally explained by the fact that 
there is a good correlation between the velocity and the intensity 
due to the thermal convection \citep[e.g.][]{1968SoPh....3..510K,
1969SoPh....9...39B} at the granular scale. Both the power spectra 
have a peak at $k\sim1$~Mm$^{-1}$, which corresponds to the spatial 
scale of about 1000~km and the granular scale. In the wavenumber range 
higher than the granular scale, i.e. in the spatial scale smaller 
than granules, the power spectra appear to be obeying the power-law 
(i. e. $\propto k^{\alpha}$). When we derive the power-law indices 
$\alpha$ of the power spectra between $k=$ 1.5 and 3.5~Mm$^{-1}$, 
the indices are -3.3 and -3.6 for $D_{th}(k)$ and $D_{vz}(k)$, 
respectively. Note that the power-law indices indicate that the 
slopes of the power spectra at the scale smaller than granules are 
significantly steeper than $k^{-5/3}$ which is predicted by the 
Kolmogorov's law for isotropic turbulences. The power-law indices 
are much steeper than the ones in \cite{1989ssg..conf..101M} and 
\cite{1993A&A...271..589E} in which they reported that the slopes 
of the power spectra are close to Kolmogorov's $k^{-5/3}$ power-law. 
Recent studies indicated power-law indices of about -3 based on a blue 
continuum image at 450~nm taken with the broadband filter imager 
(BFI) of Hinode SOT \citep{2010A&A...512A...4R} and about -4 
based on a broadband image taken through a TiO filter at 705.7~nm 
with the New Solar Telescope (NST) of Big Bear Solar Observatory 
(BBSO) \citep{2010ApJ...714L..31G}, which are not so different 
from our results.

On the other hand, the power spectra do not show any enhancements in 
the lower wavenumber domain at 0.03 -- 0.05~Mm$^{-1}$ corresponding to 
the supergranular scale. This is expected because vertical flows 
associated with supergranulation are slower than 50 m s$^{-1}$ 
\citep{2002SoPh..205...25H} on the surface, and is more than one order
of magnitude smaller than vertical flows of granulation, while horizontal
flows of supergranulation are 300 -- 500~m s$^{-1}$ 
\citep{1964ApJ...140.1120S, orozco2012} and are known to make a peak 
in power spectra at the supergranulation scale \citep{2000SoPh..193..299H, 2010A&A...512A...4R}. 

The magnetic power spectra $D_{bz}(k)$ and $D_{bh}(k)$ shown in the bottom 
panel in Fig. \ref{f1} are completely different from $D_{th}(k)$ and 
$D_{vz}(k)$. Both the horizontal and vertical powers distribute over
a broad wavenumber range. There are weak enhancements at the granular scale
($k\sim$ 1 Mm$^{-1}$) and at the supergranular scale ($k\sim$ 0.03 -- 0.05 Mm$^{-1}$),
in both $D_{bz}(k)$ and $D_{bh}(k)$, which suggests that the granular and 
supergranular scales are typical ones characterizing the magnetic field 
distribution on the solar surface. As we mentioned above, there are no 
significant powers at the supergranular scale in the spectra of the vertical 
velocities. But it is expected that the power of the horizontal velocities 
are strong enough to make the magnetic field structures, network fields, at 
the supergranular scale \citep{orozco2012}. Note that there are no clear 
enhancements of the power around the proposed mesogranular scale 
$k\sim$ 0.1 -- 0.4~Mm$^{-1}$, corresponding to 2.5 -- 10~Mm, 
\citep{1981ApJ...245L.123N,1992ApJ...396..333C} in all the power spectra 
shown in Fig. \ref{f1}, which is consistent with \cite{2000SoPh..193..299H} 
and \cite{2010A&A...512A...4R}. There are some residual magnetic powers 
in the intermediate scale between the granular and supergranular scales. 
\cite{2010ApJ...718L.171I} suggested that sporadic appearance of granular-scale 
horizontal magnetic fields preferentially happens around boundaries of 
mesogranules \citep[see also][]{2011ApJ...727L..30Y}. Such mesogranular-scale 
magnetic structures might be related with the residual magnetic power in the 
intermediate scale, though we need careful analysis including horizontal 
velocities that we cannot obtain only with the SP data. 

In the wavenumber range higher than the granular scale, the slopes of the 
magnetic power spectra are clearly less steep than the power spectra of the
thermal power $D_{th}(k)$ and the velocity power $D_{vz}(k)$. The power-law index 
$\alpha$ of $D_{bz}(k)$ is $-1.4$ in the wavenumber range between $k=$ 1.5 
and 3.5~Mm$^{-1}$ though it is hard to say that the power spectrum obeys 
the power-law only from this analysis. The horizontal magnetic power 
spectrum $D_{bh}(k)$ has its peak at 1.3 -- 2~Mm$^{-1}$ corresponding to 
500 -- 800~km, which is slightly smaller than the granular scale seen in 
$D_{vz}(k)$ and $D_{th}(k)$. The spatial scale of the magnetic spectra 
is in agreement with the observed size of transient horizontal magnetic fields 
\citep{2009ASPC..415..132I}. The slope of $D_{bh}(k)$ looks steeper than 
that of $D_{bz}(k)$ though we have to take into account relatively larger 
noises in the horizontal magnetic power. We will look into this in the next 
section.

\begin{figure*}[t]
\plotone{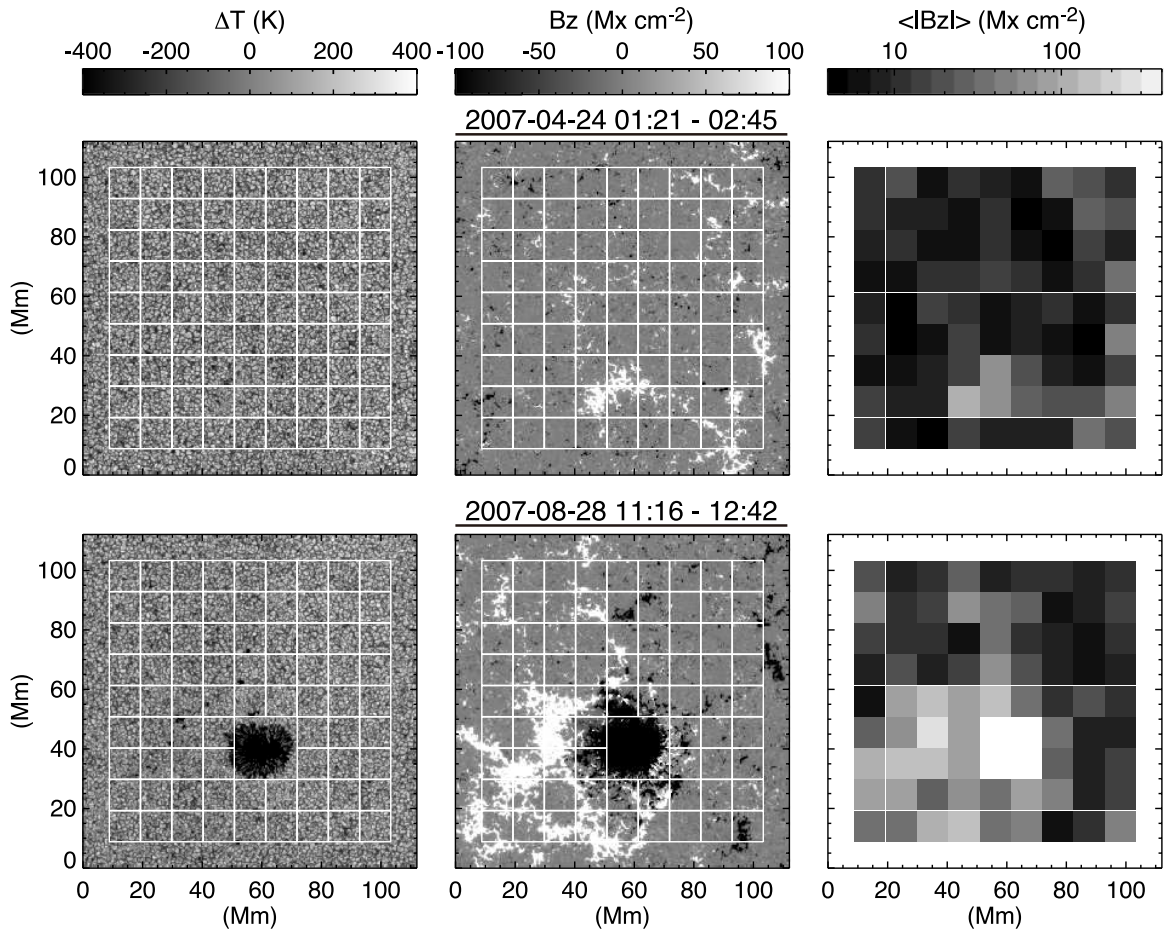}
\caption{Sub-regions used in the power spectrum analysis in Section 4 are 
shown on Hinode SP maps of the temperature fluctuation $\Delta T$ (left panels) 
and the vertical component of magnetic flux densities $B_z$ (middle panels). 
Each sub-region consists of $96\times 96$ pixels corresponding to a square 
of about 10~Mm. An average unsigned flux density $\left<|B_{z}|\right>$ is 
calculated in each sub-region as shown in the right panels. The top panels
show maps obtained on 24 April 2007 which contain only a quiet region in the
FOV. The bottom panels show maps obtained on 28 August 2007 which contain
an active region in the FOV. We do not use the regions near the boundary of 
the SP maps and sub-regions containing a sunspot in this analysis. These regions
are shown as blank areas in the $\left<|B_{z}|\right>$ maps (right panels).
\label{f3n}}
\end{figure*}

\section{Dependence of the power spectra on the unsigned magnetic flux density}

\subsection{Power spectra in small FOVs}

\begin{figure*}[t]
\plotone{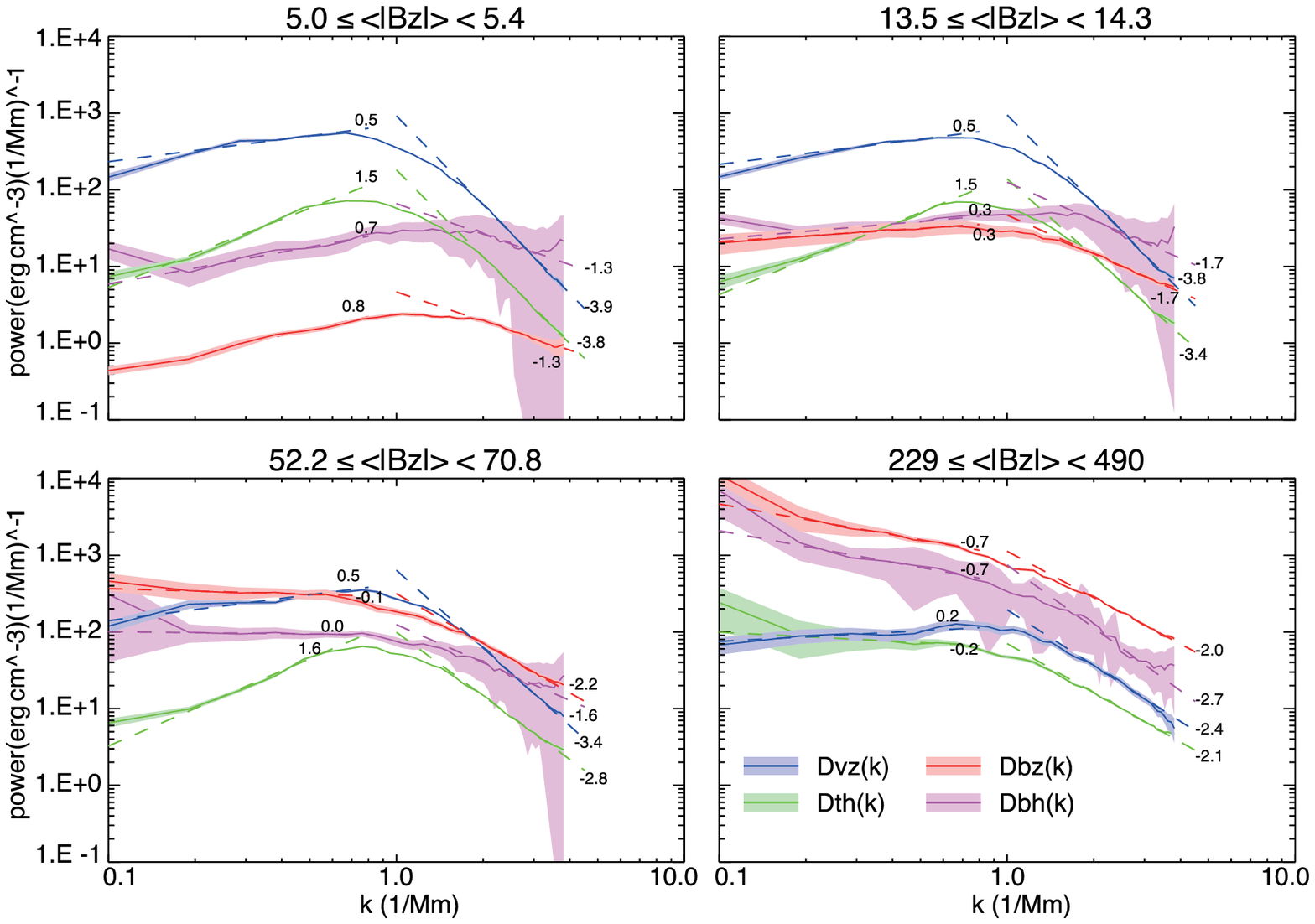}
\caption{Power spectra $D_{th}(k)$ (green curves), $D_{vz}(k)$ (blue curves), 
$D_{bz}(k)$ (red curves), and $D_{bh}(k)$ (purple curves) made in the sub-regions 
of the SP maps. Here shown are the spectra averaged in 
the intervals of 5.0~Mx~cm$^{-2}$ $\leq\left<|B_{z}|\right><$ 5.4~Mx~cm$^{-2}$
(top left), 13.5~Mx~cm$^{-2}$ $\leq\left<|B_{z}|\right><$ 14.3~Mx~cm$^{-2}$
 (top right), 52.2~Mx~cm$^{-2}$ $\leq\left<|B_{z}|\right><$ 70.8~Mx~cm$^{-2}$
  (bottom left), 229~Mx~cm$^{-2}$ $\leq\left<|B_{z}|\right><$ 490~Mx~cm$^{-2}$
 (bottom right). The shaded area of each power spectrum represents the $\pm 1\sigma$ 
 range showing variation of the power spectra in each interval of $\left<|B_{z}|\right>$. 
 The values in the plots indicate the power-law indices $\alpha$ obtained between 
 $k=$ 0.15 and 0.6~Mm$^{-1}$ and between $k=$ 1.8 and 3.2~Mm$^{-1}$.
\label{f3}}
\end{figure*}

The power spectra shown in the previous section were obtained in the quiet 
regions, thus including both internetwork and network regions. In this section, 
we analyze power spectra separately in internetwork and network regions.
To do this, each SP map is divided into $9\times 9$ sub-regions whose FOVs 
are $96\times 96$ pixels, corresponding to a square of about 10~Mm, as is shown
in Fig. \ref{f3n}. The FOVs of the sub-regions are not enough to cover the 
supergranular scale, but are enough to include many granules. The sub-regions 
are extracted not to use pixels near the edges of the SP maps. A power spectrum 
is obtained in each sub-region, using the same method described in \S 2 except 
the number of pixels in one direction $N$. The noise subtraction and the MTF 
correction are also applied to the power spectrum in each sub-region. An 
averaged density of unsigned flux $\left<|B_{z}|\right>$ is calculated in each 
sub-region, as is shown in the right panels of Fig. \ref{f3n}. The power 
spectra obtained in the sub-regions are averaged when they have similar 
$\left<|B_{z}|\right>$. The interval of $\left<|B_{z}|\right>$ for averaging the 
power spectra is determined to have 50 sub-regions in each interval. All the 30 
SP maps are used in this analysis. Some of the sub-regions include sunspots in 
their FOVs, but such sub-regions are not used by checking $\left<|B_{z}|\right>$. 

Fig.~\ref{f3} shows the thermal, kinetic, and magnetic power spectra obtained in the 
sub-regions. The power spectra are the ones averaged in the interval of 
5.0~Mx~cm$^{-2}$ $\leq\left<|B_{z}|\right><$ 5.4~Mx~cm$^{-2}$, corresponding 
to weak internetwork regions,  13.5~Mx~cm$^{-2}$ $\leq\left<|B_{z}|\right><$ 
14.3~Mx~cm$^{-2}$, corresponding to strong internetwork regions, 
52.2~Mx~cm$^{-2}$ $\leq\left<|B_{z}|\right><$ 70.8~Mx~cm$^{-2}$, 
corresponding to network regions, and 229~Mx~cm$^{-2}$ $\leq\left<|B_{z}|\right><$ 
490~Mx~cm$^{-2}$, corresponding to strong network and plage regions. 
The shaded area of each power spectrum shows variation of the power spectra
($\pm 1 \sigma$) in each interval. It is obvious that the magnetic power spectra 
of the horizontal component $D_{bh}(k)$ have large variations, which is due to 
relatively weaker signals in the linear polarization. The power spectra
are shown in the wavenumber range up to $k=3.9$~Mm$^{-1}$ because the power in
the higher wavenumber range tends to be affected by the noise subtraction.

Neither thermal or kinetic power spectra, $D_{th}(k)$ and $D_{vz}(k)$, change much 
when the average unsigned flux density is weaker than 20~Mx~cm$^{-2}$
in the internetwork regions (top two panels in Fig.~\ref{f3}). 
There are peaks at around $k\sim1$~Mm$^{-1}$ corresponding to the granular scale. 
In the wavenumber range higher than $k=1$~Mm$^{-1}$, i.e. in the spatial scale 
smaller than granules, the power-law indices are steeper than -3 in the wavenumber 
range between $k=$ 1.8 and 3.2~Mm$^{-1}$. The results are more or less consistent with the 
power spectra obtained in \S 3. On the other hand, the magnetic power spectra 
clearly varies depending on the average unsigned flux density $\left<|B_{z}|\right>$.
When the average flux density is very small (top left in Fig.~\ref{f3}), the 
shapes of the power spectra look similar between $D_{bh}(k)$ and $D_{bz}(k)$. 
Both the spectra have peaks at around $k\sim1.3$~Mm$^{-1}$ corresponding to 
about 800~km, which is slightly smaller than the granular scale and comparable 
with the observed size of horizontal magnetic field structures, as mentioned 
in the previous section. \cite{2012ApJ...746..182O} examined distribution of 
magnetic field inclinations in internetwork regions at different heliocentric 
angles, and concluded that the inclination distribution partly resulted from 
the granular-scale loop-like magnetic features. The observed peak in the 
magnetic power spectra shown in Fig. \ref{f3} probably corresponds to such 
loop-like features, which are important constituents of the internetwork 
magnetic fields. The horizontal magnetic power $D_{bh}(k)$ is also acknowledged
to be larger than the vertical one $D_{bz}(k)$ in the internetwork 
regions (top two panels in Fig.~\ref{f3}). This result looks consistent with 
\cite{2008ApJ...672.1237L} where they reported predominance of apparent 
horizontal flux in internetwork regions. The power-law indices are around $-1.3$
for both the vertical and horizontal magnetic power spectra in the wavenumber 
range between $k=$ 1.8 and 3.2~Mm$^{-1}$. Similar power-law indices of the 
magnetic power spectra were reported in previous studies 
\citep{1997SoPh..171..269L,2001SoPh..201..225A} although they were based on 
magnetic power spectra at the spatial scale larger than granules. Our magnetic 
power spectra show positive slopes at that scale in the internetwork regions.
\cite{2012A&A...541A..17S} showed that a magnetic power spectrum obtained 
with a Hinode SP data set exhibited a slope that was described by a 
power-law whose index was about $-3.2$ for spatial scales smaller than 1.2~Mm. 
The above power-law index was derived from the power spectrum without 
applying calibration of the instrument MTF. In our analysis, the raw magnetic
power spectrum before applying the MTF calibration is about $-2.7$, as is shown 
in the bottom panel of Fig.~\ref{f1}, which is not so different from the one in 
\cite{2012A&A...541A..17S}.

When the average unsigned flux density is large in the network and plage 
regions (bottom panels in Fig.~\ref{f3}), the absolute vertical magnetic power 
increases. The increase in magnetic power is especially noticeable in 
the lower wavenumber range ($k<1$~Mm$^{-1}$). The horizontal magnetic power 
also exhibits similar behavior, and increases in the lower wavenumber range.
The slope of the power $D_{bh}(k)$ and $D_{bz}(k)$ changes the sign from 
positive to negative in the wavenumber range lower than $k=1$~Mm$^{-1}$, 
which makes the peak at the granular scale less pronounced in the network 
regions. The thermal and kinetic power spectra become affected by the 
relatively strong magnetic fields in these regions. The kinetic power 
$D_{vz}(k)$ tends to decrease at the granular scale $k\sim 1$~Mm$^{-1}$, 
which is naturally explained by suppression of the convective motion by 
magnetic fields. It is also realized that the kinetic and thermal powers 
tend to be slightly enhanced in the wavenumber range higher than 2~Mm$^{-1}$.
The increase in thermal and kinetic power at the small scale can be 
attributed to contributions of faculae structures created by small-scale 
magnetic fields. The suppression of the convection at the granular scale 
and the enhancements of the small-scale power makes the power spectra 
$D_{th}(k)$ and $D_{vz}(k)$ less steep in the wavenumber range higher 
than $k=1$~Mm$^{-1}$. Surprisingly, all the thermal, kinetic, and magnetic
power spectra show similar slopes in the wavenumber range higher than 
$k=1$~Mm$^{-1}$ when the average unsigned flux density is bigger than 
200~Mx~cm$^{-2}$ (bottom right panel in Fig.~\ref{f3}).

\begin{figure}[t]
%\epsscale{0.6}
\plotone{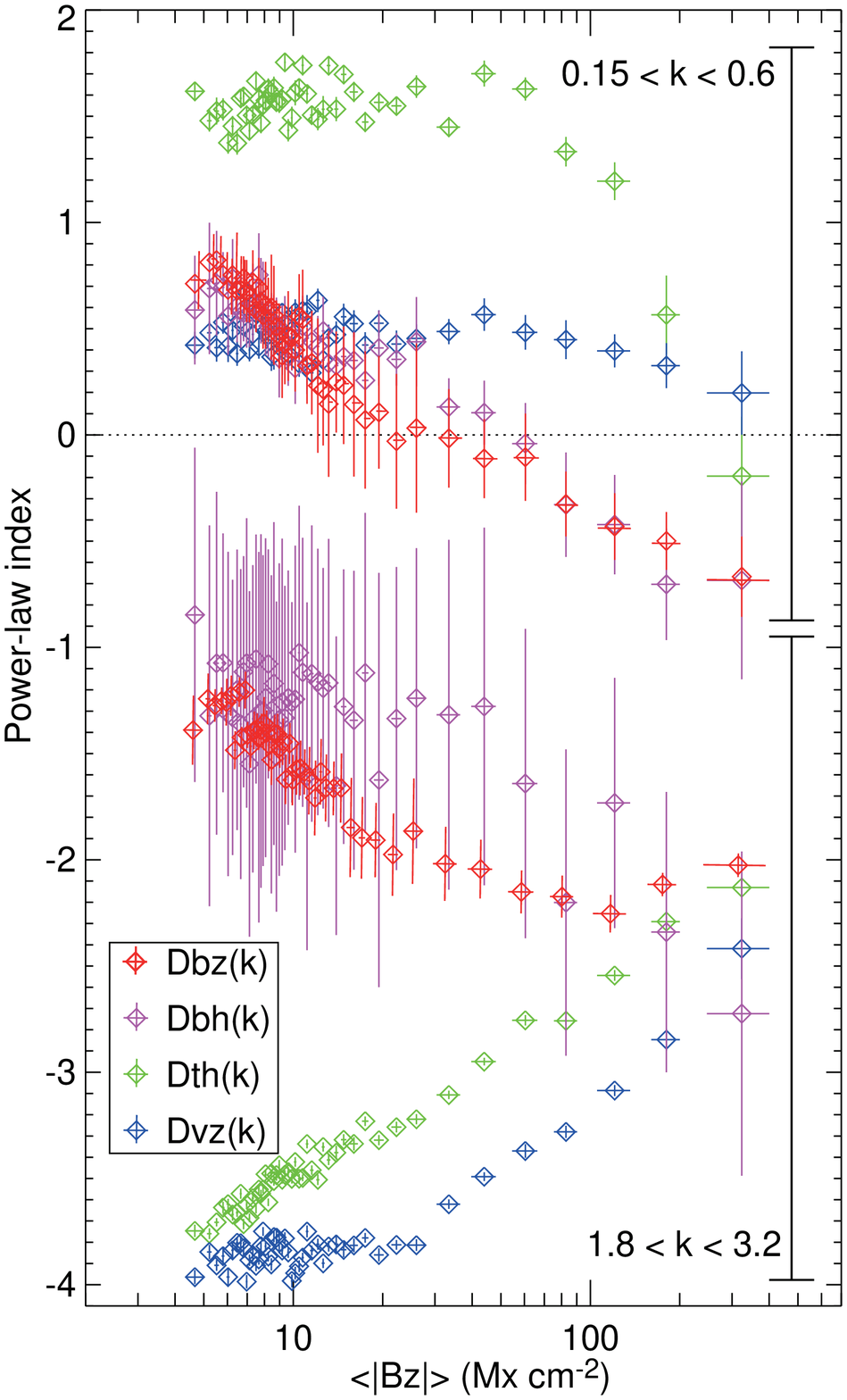}
\caption{Power-law indices $\alpha$ of the power spectra as a function of the 
averaged density of unsigned flux $\left<|B_{z}|\right>$. The power law 
indices are obtained in the two wavenumber ranges, between $k=$0.15 and 
0.6~Mm$^{-1}$ (corresponding to the scale larger than granules) and between 
$k=$1.8 and 3.2~Mm$^{-1}$ (corresponding to the scale smaller than granules). 
The green, blue, red, and purple symbols represent the power-lawn indices of 
$D_{th}(k)$, $D_{bz}(k)$, $D_{bz}(k)$, and $D_{bh}(k)$, respectively. The 
vertical error bars show $\pm1\sigma$ range of the power-law indices when 
we take into account the variation of the power spectra in each 
$\left<|B_{z}|\right>$ interval. 
\label{f4}}
\end{figure}

\subsection{Dependence of the power-law indices on the unsigned magnetic flux density}

Here we derive the power-law indices $\alpha$ of the power spectra obtained in 
the sub-regions. The power spectra used here are averaged ones 
in the interval of $\left<|B_{z}|\right>$, as described in the previous 
subsection. Thus we can get the power-law indices $\alpha$ as a function of
$\left<|B_{z}|\right>$. Because the power spectra have different slopes between 
$k<1$~Mm$^{-1}$ (larger than the granular scale) and $k>1$~Mm$^{-1}$ (smaller
than the granular scale), we derive the power-law indices $\alpha$ separately 
in these two wavenumber ranges. Here we use the wavenumber range between 
$k=$ 0.15 and 0.6~Mm$^{-1}$ and $k=$ 1.8 and 3.2~Mm$^{-1}$. We do not say that
the power spectra follow the power-law in these wavenumber ranges, but the 
indices are used to characterize the slope of the power spectra. The variation
of the power spectra in each $\left<|B_{z}|\right>$ interval is taken into 
account, and is reflected in errors of the power-law indices.

Fig.~\ref{f4} displays the power-law indices $\alpha$ as a function
of the averaged unsigned flux density $\left<|B_{z}|\right>$. Power-law indices 
of the thermal and kinetic power spectra $D_{th}(k)$ and $D_{vz}(k)$ change 
similarly as a function of $\left<|B_{z}|\right>$, while those of 
vertical and horizontal magnetic power spectra $D_{bz}(k)$ and $D_{bh}(k)$ behave 
similarly as well within the error bars. When the unsigned flux density is 
smaller than around $20$~Mx~cm$^{-2}$, the difference of the power-law 
indices is significant between the magnetic power spectra and the thermal/kinetic 
ones in the higher wavenumber range (smaller than the granular scale). The 
latter power spectra are steeper than the former ones, and the power-law indices 
differ by about 2. The magnetic and thermal/kinetic power spectra tend to have 
similar slopes when the unsigned flux density becomes large. All the 
power-law indices are around -2 when the average unsigned flux 
density is bigger than 200~Mx~cm$^{-2}$.

In the smaller wavenumber range (larger than the granular scale), the power-law
indices of the magnetic power-spectra decrease monotonically with the 
unsigned flux density $\left<|B_{z}|\right>$, as is shown in the upper part of 
Fig.~\ref{f4}. They change the sign from positive to negative when 
$\left<|B_{z}|\right>$ is 20 -- 50~Mx~cm$^{-2}$. The power-law indices of 
the thermal and kinetic power spectra do not change so much when $\left<|B_{z}|\right>$ 
is smaller than around 100~Mx~cm$^{-2}$, and they decrease when 
$\left<|B_{z}|\right>$ is bigger than around 100~Mx~cm$^{-2}$. The positive 
slopes become almost zero or negative when $\left<|B_{z}|\right>$ is around 
300~Mx~cm$^{-2}$, which means that the peaks at the granular scale disappear 
in the thermal and kinetic power spectra. When we extrapolate the trend to the 
larger flux, the power-law indices in the smaller and larger wavenumber ranges 
might become similar. This suggests that the power spectra might be represented 
as a single slope in such large magnetic fluxes, though this is out of the scope 
of this article.

\section{Summary and Discussion}

We showed the power spectral analysis of the thermal, kinetic, and magnetic 
structures on the solar surface with the data set taken with the Hinode SP,
including calibration of the instrumental effects. We showed in \S3 that 
the kinetic and thermal power spectra had a peak at the granular scale 
while the magnetic power spectra had a broadly distributed power in various
spatial scales, but had two weak peaks at both the granular and supergranular 
scales. We did not find enhancements of the magnetic power at the mesogranular
scale though there were residual powers in between the granular and supergranular
peaks. In \S4 the power spectral analysis was done in the small sub-regions
to distinguish the internetwork and network regions. We examined the slopes of 
the power spectra using the power-law indices, and compared them with the 
unsigned magnetic flux density averaged in the sub-regions. In the wavenumber 
range higher than $k=1$~Mm$^{-1}$, the thermal and kinetic power spectra 
exhibited steeper slopes than the magnetic ones in the internetwork regions,
and the power-law indices differed by about 2. The slopes of the thermal 
and kinetic power spectra became less steep when the unsigned flux density
increased, and the power-law indices of all the thermal, kinetic, and magnetic
power spectra became similar. 

In the internetwork regions, where the averaged density of the unsigned 
flux is very small, magnetic field structures are passively advected by the 
convective motion. According to \cite{CambridgeJournals:368543}, when the 
magnetic field is proportional to $(\nabla\cdot{\bf v})$ or $(\nabla\times{\bf v})$ 
in such weak field regions, its power spectrum is expected to vary as 
$D_{b}(k)\propto k^{2}D_{v}(k)$, where $D_{b}(k)$ and $D_{v}(k)$ are magnetic and 
kinetic power spectra including all the three components. This 
relationship predicts that the slope of the magnetic power spectrum is less 
steep than the kinetic one, and the difference of the power-law indices is 2, 
which is consistent with the observed difference of the power-law indices in 
the higher wavenumber range ($k>1$~Mm$^{-1}$) in the internetwork regions. 
In this analysis, we could not obtain the kinetic power spectrum in all the
three components, but only in the vertical component. We need the horizontal 
velocity components to fully justify the scenario. When the unsigned 
flux is large, the magnetic structures are no longer passive to the convective 
motion.  In this regime, magnetic and velocity fields strongly interact with each 
other, which possibly produces similar slopes between the magnetic power 
spectrum and the kinetic one. 

The total magnetic energy is provided by integration of the power spectra
in the wavenumber domain, $\int_{k_0}^{k_1} \left[D_{bz}(k)+D_{bh}(k)\right] dk$,
where $k_0$ and $k_1$ are the lower and upper ends of the wavenumber range for the 
integration. When both the power spectra are represented as a power-law 
($\propto k^{\alpha}$), the total magnetic energy is completely different in 
the case of between $\alpha<-1$ and $\alpha>-1$. When $\alpha<-1$ (steeper than 
$k^{-1}$), the total magnetic energy mainly comes from the power at the lower 
wavenumber, $D_{bz}(k_0)+D_{bh}(k_0)$. On the other hand, the total magnetic 
energy mainly comes from the power at the upper wavenumber, 
$D_{bz}(k_1)+D_{bh}(k_1)$, when $\alpha>-1$ (less steep than $k^{-1}$). 
The power-law indices of the magnetic power spectra obtained in the internetwork
regions are close to or smaller than $-1$ in the wavenumber range higher than 
$k=1$~Mm$^{-1}$. This means that the total magnetic energy mainly comes from 
either the granular-scale ($k\sim 1$~Mm$^{-1}$) magnetic structures or both the 
granular-scale and smaller ones contributing evenly.

Recent numerical simulations of the surface magneto-convection suggested that 
turbulent fields were created by small-scale dynamo action, and a significant 
amount of unsigned flux was not detected at the spatial resolution of the Hinode 
SP because of cancellation of polarization signals from opposite
magnetic polarities \citep{2007A&A...465L..43V,2009ApJ...693.1728P, 
2010ApJ...714.1606P}. These numerical simulations indicated positive slopes of 
the magnetic power spectra toward the small scales even at the spatial scale 
smaller than granules. The slopes do not agree with the magnetic power spectra 
presented in this article. If we extrapolate the observed slope $k^{-1.3}$ 
of the magnetic power spectra up to $50$~Mm$^{-1}$, an order of magnitude higher
than the Hinode SP resolution, the magnetic power in the observed wavenumber 
range between 1 and 4 Mm$^{-1}$ is about 50~\% of the power between 
1 and 50~Mm$^{-1}$, which is not as small as the percentage suggested 
by \cite{2009ApJ...693.1728P}. Because we cannot argue behavior of the power
spectra in unresolved scales based on these data, it is critical to 
use high-resolution data to extend the current analysis toward such 
small scales. \cite{2010ApJ...714L..31G} and \cite{2012ASPC..455...17A} 
showed kinetic power spectra of transverse velocities down to the 0.1~Mm 
scale based on high-resolution imaging data taken with the 1.6~meter NST 
of BBSO, but they did not have magnetic field measurements yet. One possibility
is to use magnetogram data taken with the Imaging Magnetograph eXperiment (IMaX) 
on board the 1~meter Sunrise balloon-borne observatory \citep{2011SoPh..268...57M} 
to extend the power spectra down to the 0.1~Mm scale. \cite{2012A&A...541A..17S} 
argued possible existence of a flux tube population at the spatial scale close 
to or smaller than the resolution limit of the Hinode SP using a deep mode
data set. Because the subtraction of the noise spectrum becomes critical to 
derive the power in the higher wavenumber range, we could not argue the power at 
the spatial scale below 0.25~Mm using the normal mode data in this article. It is 
important to use polarimetric data with a higher signal-to-noise ratio for studying 
the power at the spatial scale as close as the instrumental resolution limit.

\acknowledgements
The authors would like to thank an anonymous referee for his/her valuable 
comments and suggestions. They also acknowledge Y. Suematsu for providing the
point spread function of the Hinode SP. This research was funded in part by 
the Ito Science Foundation. D. O. S. thanks financial support by the Spanish 
Ministry of Economy and Competitiveness (MINECO) through the project 
AYA2010-18029 (Solar Magnetism and Astrophysical Spectropolarimetry). 
HINODE is a Japanese mission developed and launched by ISAS/JAXA, with NAOJ 
as domestic partner and NASA and STFC (UK) as international partners. It is 
operated by these agencies in co-operation with ESA and NSC (Norway).

\end{document}